\documentclass[prl,amsmath,amssymb,nofootinbib,showpacs,twocolumn]{revtex4-2}
\usepackage{overpic}
\usepackage{color}
\usepackage{graphicx}
\usepackage{amsmath,amssymb,epsfig}
\usepackage{tikz}
\tikzset{box1/.style={draw=grey, thin, rectangle, minimum height=2.1cm, minimum width=3.5cm}}

\newcommand{\dd}{\partial}
\newcommand{\rd}{\mathrm{d}}

\newcommand{\td}[2]{\frac{\rd #1}{\rd #2}}

\newcommand{\eps}{\varepsilon}

\newcommand{\beq}{\begin{equation}}
\newcommand{\eeq}{\end{equation}}

\usepackage{cleveref}
\usepackage{transparent}

\crefname{equation}{Eq.}{Eqs.}
\Crefname{equation}{Equation}{Equations}
\Crefname{figure}{Fig.}{Figs.}
\crefname{section}{\S}{\S\S}

\newcommand{\orcid}[1]{\href{https://orcid.org/#1}{\includegraphics[width=8pt]{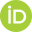}}}

\begin{document}

\title{Dynamically triggered snap-through}
\author{Gregory Kozyreff\,\orcid{0000-0002-7126-1633},$^{a}$ Lucie Domino\,\orcid{0000-0002-8064-2075},$^{b}$ Basile Radisson\,\orcid{0000-0002-0661-245X},$^{c}$  Hadrien Bense\,\orcid{0000-0003-2498-422X}\,$^{d}$}
\affiliation{
$^{a}$D\'epartement de Physique, Facult\'e des Sciences, Universit\'e libre de Bruxelles (U.L.B.), CP 231, Belgium 
\\
$^{b}$Aix Marseille University, CNRS, Institut Universitaire des Syst\`{e}mes Thermiques et Industriels, Marseille 13453, France 
\\
$^{c}$Sorbonne Université, CNRS, Institut Jean Le Rond d’Alembert, F-75005 Paris, France
\\
$^{d}$Laboratoire d'Acoustique de l'Universit\'{e} du Mans (LAUM), IA-GS, CNRS, 72085 Le Mans, France
}

\begin{abstract}

Snap-through is a sudden and large mechanical deformation that is usually triggered by driving a control parameter beyond a bifurcation threshold. Here, we study a buckled strip that undergoes a saddle-node bifurcation as one of its clamps is rotated. We show experimentally, theoretically, and numerically that the effective snapping threshold can be markedly lowered by speeding up the rotation. For actuation times that are short but nevertheless 
longer than  the fundamental flexural period, the strip  is observed to remain close to the instantaneous quasi-static equilibrium for most of the loading process. This allows us to derive an energetic criterion that accurately estimates a dynamic instability threshold well below the saddle-node bifurcation point. On the other hand, very near the bifurcation point, the evolution of the quasi-static equilibrium undergoes a geometric boost, which entrains the general motion of the beam. As a result, a slower actuation of the clamp than predicted by the energy argument suffices to produce snap-through. A distinct scaling is derived in that limit, based on the aforementioned boost and critical slowing down of the least stable vibration mode of the beam. More broadly, our results show how finite-rate forcing can trigger snap-through before the quasi-static instability threshold is reached.

\end{abstract}
\date{\today}
\maketitle

\begin{figure}
    \centering
    \includegraphics[width=\linewidth]{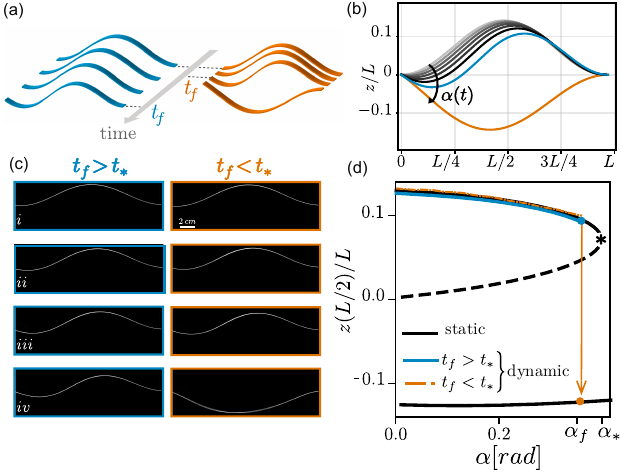}
    \caption{
    a) A buckled strip of length $L$ is actuated so that its edge takes an angle $\alpha_f$; depending on the speed of actuation $\alpha_f/t_f$, the strip might snap through (orange strips) or not (blue strips). 
    b) Side view of the strips, showing the evolution of the shape with $\alpha(t)$ and the two final configurations. 
    c) Chronophotographs of a buckled strip actuated from $\alpha = 0\,$rad to $\alpha_f < \alpha_*$. Left column: the actuation time $t_f>t_*$, the strip does not snap-through. Right column: $t_f<t_*$, snap-through is triggered. The times for the photographs are $t/t_f$ 0, 0.5, 1 and 4.
    d) Bifurcation diagram of a buckled strip under a strain of $5\%$. Black lines show the static shapes for the first two stable (plain lines) and unstable (dashed line) solutions. Quasi-static snap-through occurs when the unstable and the stable solution meet (star symbol). In this Letter we consider dynamic actuations that stop at $\alpha_f<\alpha_*$. Depending on the speed of the actuation, the strip can snap through (orange line) or not (blue line).  
    }
    \label{fig:1}
\end{figure}

Mechanical instabilities of  slender elastic bodies have extended beyond the exclusive domain of engineering and have become in recent years a valuable theoretical guide to study a variety of phenomena in nature and to design smart structures~\cite{reis2015}. Snap-through buckling is a particular form of mechanical instability in which a multistable system suddenly jumps from one stable state to another, and which can be harnessed in numerous systems. Venus flytraps~\cite{Forterre2005,skotheim2005} and hummingbirds~\cite{smith2011} utilise the instability to generate rapid predatory movements; cicadas produce sound via the snap-through of tymbal membranes in their abdomen~\cite{ghoshal2025}. The nonlinear behaviour of bistable structures is increasingly exploited in the design of functional devices, be they  actuators~\cite{gude2011,ducarme2025}, energy harvesters~\cite{harne2013} or absorbers~\cite{shan2015}, MEMS~\cite{wang2013,Cao2021}, robotics~\cite{Wang2023_pnas} or flexible mechanical metamaterials~\cite{Sen2025,Deng2021,Librandi2021,liu2024}. Beyond their technological relevance, such instabilities also provide simple model systems for studying nonlinear dynamics close to bifurcation points. In that limit, the time scale associated to the lowest mode of vibration diverges -- a phenomenon called ``critical slowing down''. As a result, the jump towards the new mechanical state of deformation can be much slower than what is expected from the geometrical and elastic constants of the system~\cite{Gomez2017}. By the same token, sweeping the control parameter across the bifurcation threshold at the appropriate speed, one can either delay~\cite{Liu2021,Huang2024} or control the route to snap-through in the phase plane~\cite{Wang2024}. Note also that the nature of the bifurcation itself--saddle-node or subcritical pitchfork--depends on the symmetry of the loading condition, with incidence on the time scales involved~\cite{Radisson2023a,Radisson2023b}. In these studies, the control parameter is always brought beyond the instability threshold. Here, we address a different question: can one induce bifurcation by sweeping the control parameter at finite speed without crossing the snapping threshold? To this end, we focus on a model experiment of a buckled beam where one end is rotated at constant angular speed over a finite actuation time (\Cref{fig:1}a). The final angle is always below the static critical angle for snapping (\Cref{fig:1}b). Combining experiments, theory and numerical simulations, we determine the dynamic instability threshold and identify the physical mechanisms governing its dependence on the actuation time (\Cref{fig:1}c).

\textit{Experiments}--A flat, horizontal strip of length $L$ is initially made to buckle sideways by  compressing it over a distance $\Delta L=\eps L$, where $\eps$ is the compressive strain. Initially, upwards or downwards buckling are symmetric and energetically equivalent. The symmetry, however, is then broken by the rotation of one of the clamps through an angle $\alpha$ (\Cref{fig:1}b). At a critical angle $\alpha_*$, the more bent of the two stable solutions merges with an unstable one and disappears through a saddle-node bifurcation (\Cref{fig:1}c).
As pointed out by Gomez \textit{et al.} \cite{Gomez2017},  the proper control parameter is
\beq
\mu=\alpha /\sqrt{\eps}
\eeq
and modelling the strip using the classical Euler beam equation yields the static instability threshold  $\mu_*\approx1.78$.  In our experiments, we vary $\alpha$, keeping $\eps$ fixed; we will therefore find it convenient to discuss our results mostly in terms of  $\alpha$, bearing in mind that it is $\mu$ that ultimately matters. Intuitively, if we let $\alpha$ increase linearly in time from $0$ up to a final value $\alpha_f$ in a finite actuation time $t_f$, it should be possible to induce snap-through even if $\alpha_f<\alpha_*=\sqrt{\eps} \mu_*$ (\Cref{fig:1}a). 

To verify this hypothesis, we carried out experiments on thin strips of polyethylene terephtalate (PET) with  Young's modulus  $E\approx 5\times 10^9\,Pa$, Poisson's ratio $\nu \approx 0.4$, length $L$ in the range $10 - 20\,\text{cm}$, width $b\approx 6\,\text{mm}$ and a thickness $h$ in the range $100- 250\mu \text{m}$, so that $L \gg b\gg h$. The material density is $\rho=1380\,kg\,m^{-3}$ and we denote the mass per unit length $\rho b h$ by $\rho_l$. We remain in the limit of small strain with $\varepsilon\leq5\%$. One end of the strip is connected to a stepper motor (RS Pro) which allows us to control the clamp angle, and its angular speed. The experiment is monitored with a high speed camera at a frame rate of about $500\,$fps. For given strip characteristics and target angle $\alpha_f$, we impose linear ramps of clamp rotation
\beq
\alpha(t)=\alpha_{f}\times\left\{
\begin{matrix}0,& t<0,\\  \min(t/t_f,1),& t>0,\end{matrix}
\right.
\label{eq:BC}
\eeq
and decrease $t_f$ until we find a critical value $t_*$ of the actuation time. For $t_f>t_*$, the strip does not snap-through: at the end of the actuation, the strip oscillates around its initial buckled position and remains in this configuration (Fig.~\ref{fig:1}a and c left row). Conversely, for $t_f<t_*$ snap-through is observed (Fig.~\ref{fig:1}a and c right row). We measure $t_*$ for a range of target angles $\alpha_f$, strip geometries and compressive strains.
\begin{figure}
    \centering
    \includegraphics[width=\linewidth]{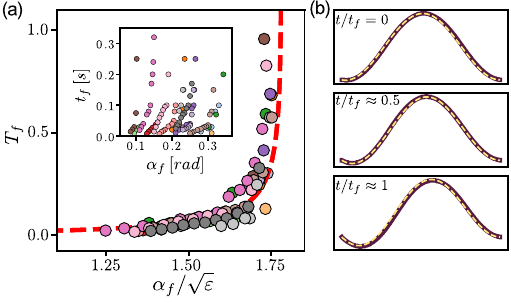}
\caption{(a) Experimental, rescaled,
actuation time $T_f = t_f/t_{el}$ 
below which snapping is triggered as a function of the rescaled final clamp angle $\mu_f = \alpha_f/\sqrt{\eps}$
under various compressions, beam lengths and thicknesses~\cite{supp}. 
Red dashed line: theoretical curve based on \cref{eq:kinetic,eq:barrier}, without a fitting parameter. Inset: raw experimental data $t_f$ as a function of $\alpha_f$. 
(b) Purple: experimental snapshots of the strip during actuation ($t_f=0.04\,$s, $\alpha_f=0.26\,$rad, $\eps\approx 5\,\%$); yellow: static solution of Euler's beam equation at the same angle.
}
\label{fig:exp}
\end{figure} 
As expected, $t_*$ increases with $\alpha_f$ and diverges as $\alpha_f\to\alpha_*$ (inset of \Cref{fig:exp}a). Qualitatively, the amount of kinetic energy that must be imparted to the strip to trigger snapping decreases with distance from the instability, with the limit case of a zero speed when $\alpha_f \to \alpha_*$. A natural timescale for this problem is well known to be $t_{el}=\sqrt{\rho_lL^4/B}$, where $B = \frac{Eh^3b}{12(1-\nu^2)}$ is the bending rigidity~\cite{Gomez2017,Radisson2023a}. Rescaling our data as $T_f=t_f/t_{el}$ and plotting them as a function of $\mu_f=\alpha_f/\sqrt{\eps}$, the experimental points collapse on a single master curve (\Cref{fig:exp}a). This completes the experimental demonstration that, for a given final angle $\alpha_f<\alpha_*$, a well-defined critical actuation time $t_*$ exists below which snap-through occurs.

\textit{Quasi-static regime ---} High-speed recordings show that, throughout the actuation, the beam deformation remains remarkably close to the corresponding static equilibrium profile at the same instantaneous clamp angle (\Cref{fig:exp}b). This observation suggests that, in the experimentally tested range of actuation, the dynamic threshold is not governed by the detailed transient deformation of the beam, but rather by the kinetic energy accumulated while the structure follows the equilibrium branch. In this frame, we make two working hypotheses. Firstly (H1), following experimental evidence, we assume that the strip deflection remains close to the static solution of the Euler beam equation, $z_s(x,\alpha)$, during the loading phase (top branch in \Cref{fig:1} d). Secondly (H2), we assume that the local vertical speed of the strip is approximately constant during the loading phase. Then, the kinetic energy imparted to the structure can be estimated as:
\beq
U_K
=\frac{\rho_l}{2t_f^2}\int_0^L\left[z_s(x,\alpha_f)-z_s(x,0)\right]^2\rd x.
\label{eq:kinetic}
\eeq

On the other hand, the energy barrier to snap-through is simply the difference in elastic energy between the shape at the end of the actuation $z_s(x,\alpha_f)$ and the unstable shape $z_u(x,\alpha_f)$ at the same angle. (dotted branch in \Cref{fig:1}d) (stretching energy is negligible due to the slenderness of the strip). It reads:
\beq
U_e
=\frac{B}{2}\int_0^L\left\{\left[\dd_{x}^2z_u(x,\alpha_f) \right]^2-\left[\dd_{x}^2 z_s(x,\alpha_f) \right]^2\right\}\rd x,
\label{eq:barrier}
\eeq
where $\dd_x$ denotes differentiation with respect to $x$. 
We posit that snap-through occurs when $t_f=t_*$, such that $U_K = U_e$, allowing us to estimate $t_*$ from the sole knowledge of $z_s(x,0)$, $z_s(x,\alpha_f)$ and $z_u(x,\alpha_f)$. After rescaling, this yields the dashed red line in \Cref{fig:exp}b. This theoretical curve, without any fitting parameter, agrees well  with the experimental data, thereby validating our energy argument. We expect, however, that the  hypotheses behind the foregoing theory are not valid for any actuation speed. A finer understanding requires us to study in more detail the deformation of the structure.

\textit{Slow variation of $\alpha$ away from the saddle-node bifurcation}---
For actuation times longer than the period of the first flexural mode, two well-separated time scales coexist: the slow evolution imposed by the clamp rotation and the fast oscillations of small amplitudes. These can be handled by a WKB approach~\cite{Kozyreff2026}. We denote $z(x,t)$ the transverse displacement of the beam and we adopt $\tau = t/t_f$ as the time variable. The  beam equation then reads~\cite{Howell2009}: 
\beq
t_f^{-2}\,\rho_l\,\dd_\tau^2 z+B\,\dd_x^4z+P(\tau)\, \dd_x^2z=0,
\label{eq:beam}
\eeq
where $\dd_\tau$ denotes partial differentiation with respect to $\tau$. $P(\tau)$ is the compressive load, determined by the integral constraint
\beq
\frac{1}{L}\int_0^L\left(\dd_x z\right)^2\rd x = 2\eps,
\label{eq:NL}
\eeq
which ensures the constancy of the length of the strip. \Cref{eq:NL} turns the differential problem into a nonlinear one, with multiple static solutions (\Cref{fig:1}d). It is completed by the boundary conditions $z(0,\tau)=z(L,\tau)= \dd_xz(L,\tau)=0$ and $\dd_x z(0,\tau)=\alpha(\tau)$. In the limit of long actuation times, \textit{i.e.} $t_{el}/t_f\ll1$, $t_f^{-1}$ is the small parameter of the WKB expansion. Let us first decompose both the displacement of the strip and the compressive force as the sum of a quasistatic part and a small deviation from it:
\begin{align}
z(x,\tau)&=z_s\left(x,\alpha(\tau)\right)
+ \operatorname{Re}\left[e^{it_fS(\tau)}\Phi(x,\tau)\right],
\label{WKB:z}
\\
P(\tau)&=P_s\left(\alpha(\tau)\right)
+ \operatorname{Re}\left[e^{it_fS(\tau)}\Psi(\tau)\right].
\label{WKB:P}
\end{align}
with $\Phi(0,\tau) = \Phi(L,\tau) = \dd_x\Phi(0,\tau)=\dd_x\Phi(L,\tau)=0$ and $\Psi(0)=0$. Neglecting the product $\Psi\dd_x^2\Phi$, \cref{eq:beam} becomes 
\begin{multline}
t_f^{-2}\partial_\tau^2 z_s
+\operatorname{Re}\Big \{
\Big[
-S'^2\Phi
+2it_f^{-1}S'\partial_\tau\Phi
+it_f^{-1}S''\Phi
+t_f^{-2}\partial_\tau^2\Phi
\\
+\frac{B}{\rho_l}\partial_x^4\Phi
+\frac{P_s}{\rho_l}\partial_x^2\Phi
+\frac{\Psi}{\rho_l}\partial_x^2 z_s
\Big] e^{it_fS(\tau)}
\Big \}
=0,
\label{eq:MMS1}
\end{multline}
where $S' = \frac{dS}{d\tau}$.
We first set  $S'(\tau) = \omega_n(\tau)$, where $\omega_n(\tau)$ is the instantaneous angular frequency associated to the $n^{th}$ linear mode of vibration around $z_s(x,\alpha(\tau))$. Then, expanding  $\Phi$ as $\Phi \sim \Phi_0+t_f^{-1}\Phi_1+\dotsb$ and similarly for $\Psi$, we get, at leading order in $t_f^{-1}$,
\begin{equation}
\rho_l\omega_n^2\Phi_0 + B\dd_x^4\Phi_0+P_s\dd_x^2\Phi_0+\Psi_0\dd_x^2z_s =0,
\label{freevib}
\end{equation}
while the nonlinear constraint yields $\int_0^L\dd_xz_s\dd_x\Phi_0dx=0$. The problem just derived for $\Phi_0$ and $\Psi_0$ coincides with that of deriving the free vibrations of a beam clamped with an angle $\alpha(\tau)$, in which $\tau$ only appears as a parameter. Following \cite{Pandey2014,Radisson2023b} we have
\begin{align}
\begin{pmatrix}\Phi_0\\ \Psi_0\end{pmatrix}&=A_n(\tau)
\begin{pmatrix}w_n(x,\alpha(\tau))\\ p_n(\tau)\end{pmatrix},
\end{align}
where $w_n$ and $p_n$ are respectively the $n^{th}$ resonant flexural mode and compressive force associated to $\alpha(\tau)$. The evolution of the amplitude $A_n(\tau)$ is determined by going to the next order and writing the solvability condition. Provided that the vibration modes are normalised, $\int_0^Lw_n^2\rd x = 1$, we derive \cite{supp}:
\begin{equation}
    A_n(\tau) = A_n(0)\sqrt{\frac{\omega_n(0)}{\omega_n(\tau)}}.
\end{equation}
Eventually, taking all modes into account and reverting to the original time variable, we obtain
\begin{multline}
z(x,t) \sim z_s(x,\alpha(t)) + 
\\
\sum_n \operatorname{Re}\left[A_n(0)\sqrt{\frac{\omega_n(0)}{\omega_n(t)}}e^{i\int_0^t\omega_n(t')\rd t'}w_n(x,\alpha(t))\right].
\label{eq:modes}
\end{multline}
This formula holds as long as the actuation time is long compared to the period of the fundamental flexural mode. Initially, $\omega_0(0)/2\pi\approx7/t_{el}$~\cite{Radisson2023b}, so it is only required that $t_f>t_{el}/7$. Furthermore, for $\alpha(\tau)$ given by \cref{eq:BC}, we find that  $A_n(0)\propto \alpha_f/[\omega_n(0)t_f]=\mu_f\sqrt{\eps}/[\omega_n(0)t_f]\ll1$. The asymptotic analysis thus provides a clear justification of the empirical quasi-static picture introduced above (\Cref{fig:exp}b). In practice, the small amplitude oscillations are difficult to measure experimentally. Therefore, in the next part, we use numerical simulations to validate the asymptotic structure of the solution.

\begin{figure}
    \centering
    \includegraphics[width=.9\linewidth]{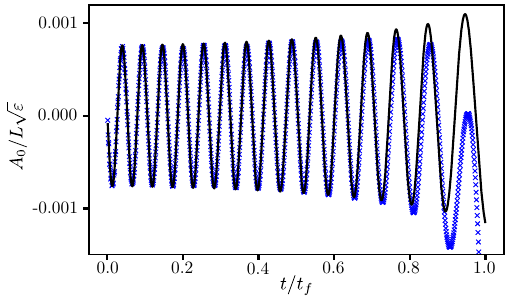}
    \caption{Normalised coefficient of the projection on the first dynamical mode $A_0/L\sqrt{\varepsilon}$ as a function of $t/t_f$. $\mu_f=1.72$ and $t_f = 0.125\,$s, $t_{el}\approx 0.0448\,$s. Blue symbols are obtained from numerical simulations, the black line is calculated from \cref{eq:modes}. No adjustable parameters.}
    \label{fig:MMS}
\end{figure}

\textit{Numerical results}---
We simulate the motion of the strip using the discrete Cosserat rod formulation~\cite{Gazzola2018}.  Within this numerical scheme the actual critical point is found to be $\mu_*\approx 1.75$,  which is slightly less than the value predicted by the Euler beam model. In the simulations, $\eps$ is kept constant while $\alpha$ is varied at different angular speeds, effectively bringing $\mu$ from $0$ to $\mu_f = 1.72$. We project the numerically computed difference $z(x,t)-z_s(x,\alpha(t))$ on the modes $w_n(x,\alpha(t))$ in order to extract their associated amplitudes $A_n(t)$. The agreement  with the WKB approximation $A_n(0)\sqrt{\frac{\omega_n(0)}{\omega_n(t)}}e^{i\int_0^t\omega_n(t')dt'}$ is excellent, with no adjustable parameters  (\Cref{fig:MMS}). The first and second modes concentrate most of the oscillations. 
We note that the amplitude $A_0(t)$ increases over time, because $\omega_0(t)\to 0$ as $\alpha(t)\to \alpha_*$. Close to the saddle-node bifurcation, \textit{i.e}, $t/t_f\approx 1$ the  numerical simulation starts to depart from the WKB approximation: the actual dynamic shape diverges from the static one (Fig.~\ref{fig:MMS}). Indeed, in the vicinity of $\alpha_*$, the stable branch of static solutions varies more and more rapidly with $\alpha$ (\Cref{fig:1}d), so that hypothesis H2 behind the quasi-static approximation progressively breaks down and the strip can no longer follow it adiabatically. Consequently,  the WKB asymptotic structure of the solution is no longer accurate and the simple energetic criterion at the beginning of our analysis ceases to correctly predict the critical actuation time. Furthermore, nonlinear effects can no longer be neglected. The WKB approximation must therefore be revised.

\begin{figure}
    \centering
    \includegraphics[width=\linewidth]{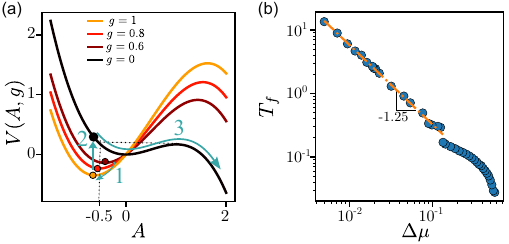}
    \caption{(a) Evolution of the potential $V(A,g)$. 1/ During slow actuation, the amplitude adiabatically follows the minimum of the potential (carmine to yellow dots). 2/ When actuation ceases, $g$ suddenly vanishes, hence $V$ increases. 3/ The amplitude evolves freely, if $V(A(t_f),g=0)$ (black dot) is high enough, the beam has enough energy to cross the unstable point, with unbounded growth of $A$. b) $T_*$ obtained numerically as a function of $\Delta \mu = \mu_*-\mu_f$ in a log-log representation. Near the saddle-node bifurcation, we observe that $T_* \sim \Delta \mu^{-1.247}\approx \Delta\mu^{-5/4}$, as predicted theoretically.}
    
    \label{fig:4}
\end{figure}

\textit{Close to the saddle-node bifurcation---} The fundamental mode of vibration $w_0(x,\alpha)$ prevails over the others and its associated frequency decreases as $\omega_0=O[(\alpha_*-\alpha)^{1/4}]$ (see \cite{Radisson2023b}). This invites us to recast \cref{WKB:z,WKB:P} as
\begin{align}
z(x,t)&\sim z_s(x,\alpha(t)) + A(t)w_0(x,\alpha(t))+\delta\zeta,\\
P(t)&\sim P_s(\alpha(t))+A(t)p_0(\alpha(t))+\delta p,
\end{align}
where $A(t)$ is a slow-varying amplitude and $\delta\zeta$ and $\delta p$ are second-order deviations. Crucially,  $z_s(x,\alpha(t))$ differs from its limiting value $z_s(x,\alpha_*)$ only by an $O[(\alpha_*-\alpha(t))^{1/2}]$ amount (\Cref{fig:1}d). As a result, $\dd_t^2z_s(x,\alpha(t))$, which used to be negligible in \cref{eq:beam}, is now boosted by an  $O[(\alpha_*-\alpha(t))^{-3/2}]$ factor. In addition to this, the strip is accelerated by $A''(t)w_0$ and is subjected to a local restoring force  $\rho_l\omega_0^2Aw_0$,  together with a weakly nonlinear force  due to the proximity of the saddle-node bifurcation point. Bringing all these terms in balance  and applying a solvability condition \cite{supp} we derive:
\beq
\td{^2A}{t^2}+\omega_0(t)^2A+g(t) = \gamma A^2,
 \label{eq:amplitude}
\eeq
where
\beq
g(t) = \frac{\int_0^Lw_0\dd_t^2 z_sdx}{\int_0^Lw_0^2dx}, \quad \gamma = \frac{3p_0}{2\rho_l}\frac{\int_0^L(\dd_x w_0)^2dx}{\int_0^Lw_0^2dx}.
\eeq
\Cref{eq:amplitude} captures the essential dynamics of the beam near $\alpha_*$. Its left-hand side is identical to the equation of a harmonic oscillator in an accelerated frame: $g(t)$ is the acceleration of the suitably averaged reference state towards which the beam is drawn, with an effective spring constant $\propto \omega_0^2$. It is interesting to compare \cref{eq:amplitude} with $\td{^2A}{t^2} = \gamma A^2$, which was derived by Gomez \textit{et al.} for a similar mechanical set-up~\cite{Gomez2017}. In their study, the clamp was fixed ($g=0$) and the beam was prepared in a deformed state \textit{beyond} the saddle-node bifurcation, so that there was no restoring force towards that initial condition ($\omega_0=0$). Without embarking on a complete study of \cref{eq:amplitude}, one can deduce a simple scenario for snapping as follows. Assume that $g(t),\omega_0(t)$ both vary sufficiently slowly that $A''(t)$ can initially be neglected in \cref{eq:amplitude}. Then, if $A$ is sufficiently small, $A(t)\approx -g(t)/\omega_0^2(t)$ for $0<t<t_f$. More precisely, $A(t)$ adiabatically follows the local minimum of $V(A,g) = gA+\frac{\omega_0^2}{2}A^2 -\frac{\gamma}{3}A^3$ (red to yellow dots in \Cref{fig:4}a). At $t=t_f$, actuation ceases and $g$ suddenly vanishes, causing a sudden jump of potential (yellow to black dot in fig.~\ref{fig:4}a). From this point on, $A(t)$ evolves in a constant potential, with the initial value  $A(t_f)\approx -g(t_f)/\omega_0(t_f)^2$, and speed $A'(t_f)=v_f$. The latter $v_f$ arises from a Dirac delta function in $g(t)$ at $t=t_f$ \cite{supp}. Since \cref{eq:amplitude} is autonomous for $t> t_f$, an elementary classical mechanical analysis becomes applicable. In particular, we can see in \Cref{fig:4}a that the actuation history prior to $t_f$ in conjunction with the jump of potential can be such that $A(t)$ overcomes the potential barrier for $t>t_f$, eventually leading to snapping.

To close the analysis, let us examine the scalings that allow each term in \cref{eq:amplitude} to balance. Given that $\omega_0(t_f)=O[(\alpha_*-\alpha_f)^{1/4}]$, balancing $\gamma A^2$ with $\omega_0^2A$ yields $A=O[(\alpha_*-\alpha_f)^{1/2}]$ and, hence, $\omega_0^2A=O(\alpha_*-\alpha)$. Next, $g(t)=O[(\alpha_*-\alpha_f)^{-3/2}t_f^{-2}]$. Hence, $g$ and $\omega_0^2A$ balance if $t_f=O[(\alpha_*-\alpha_f)^{-5/4}]$. In proper dimensionless form, the latter condition is $T_f=O[(\mu_*-\mu_f)^{-5/4}]$. The critical actuation time for snapping must therefore follow this scaling, which we confirm numerically in \Cref{fig:4}. We note that it is consistent with the one recently derived by Huang \textit{et al.}, who observed delayed snap-through by sweeping the actuation angle \textit{beyond} $\alpha_*$~\cite{Huang2024}.

In this Letter, we have studied the effect of a finite actuation time on snapping.
First, we have identified a regime where  simple energy arguments are enough to reliably determine the conditions for snapping. The actuation must not be too fast, otherwise a large number of modes can be excited with finite amplitude, leading to complex spatio-temporal dynamics that escape simple analysis. More surprising is the analysis near the saddle-node bifurcation. The non-analytic dependence of the static beam shape on the control parameter boosts the acceleration of the reference state. Combined with critical slowing down, this acceleration has a catapulting effect on the least stable mode of oscillation. The consequence of this process complements the conclusion of Gomez \textit{et al.}~\cite{Gomez2017} in an interesting way: while it is true that the snapping dynamics is slowed down by the proximity of the saddle-node point, it is also made considerably easier to trigger than what simple energy arguments suggest. The structure is therefore more vulnerable to snapping, an observation that could be important in applied contexts.

\bibliography{elastic}

\section*{Acknowledgements}
 HB acknowledges funding from the European Union’s Horizon 2020 research and innovation programme under the Marie Sklodowska-Curie grant agreement number 101102728. HB also acknowledges funding from PULSAR - L'Académie des jeunes chercheurs en Pays de la Loire. GK was supported by the Fonds de la Recherche Scientifique -FNRS. BR acknowledges funding from the European Union’s Horizon Europe research and innovation programme under the Marie Sklodowska-Curie grant agreement number 101205621.




\end{document}